\documentclass[10pt,english]{revtex4}
\usepackage{amssymb,amsmath,amscd}

\begin{document}

\title{Spin connection as Lorentz gauge field in Fairchild's action.}

\author{Francesco Cianfrani}
\affiliation{Institute for Theoretical Physics, University of Wroc\l{}aw, Pl.\ Maksa Borna 9, Pl--50-204 Wroc\l{}aw, Poland.}

\author{Giovanni Montani}
\affiliation{Dipartimento di Fisica, Universit\`a degli studi di Roma "La Sapienza", P.le A. Moro 5 (00185) Roma, Italy.\\
ENEA, Unit\`a Tecnica Fusione, ENEA C. R. Frascati, via E. Fermi 45, 00044 Frascati (Roma), Italy.
}

\author{Vincenzo Scopelliti}
\affiliation{Dipartimento di Fisica, Universit\`a degli studi di Roma "La Sapienza", P.le A. Moro 5 (00185) Roma, Italy.\\
Institute Lorentz for Theoretical Physics, Leiden University, Leiden 2300RA, The Netherlands.}





\begin{abstract}
We propose a modified gravitational action containing besides the Einstein-Cartan term some quadratic contributions resembling the Yang-Mills lagrangian for the Lorentz spin connections. We outline how a propagating torsion arises and we solve explicitly the linearised equations of motion on a Minkowski background. We identify among torsion components six degrees of freedom: one is carried by a pseudo-scalar particle, five by a tachyon field. By adding spinor fields and neglecting backreaction on the geometry, we point out how only the pseudo-scalar particle couples directly with fermions, but the resulting coupling constant is suppressed by the ratio between fermion and Planck masses. Including backreaction, we demonstrate how the tachyon field provides causality violation in the matter sector, via an interaction mediated by gravitational waves. 

\end{abstract}





\maketitle

\section{Introduction}

As pointed out by the Einstein-Cartan theory \cite{Cartan1,Cartan, Kibble, Sciama}, it is  possible to implement a local symmetry in the description of the space-time  by using the tetrad formalism. In such a formulation, a local basis of tangent space is introduced  and the theory is characterised by a local Lorentz rotation symmetry.

In his seminal paper \cite{Utiyama}, Utyiama proposed a method to introduce gauge fields associated with Lorentz transformations; he showed that these fields are nothing more than the spin connections $\omega_{\mu}^{IJ}$ corresponding to a physical gauge symmetry as the space-time description is unaffected by a Lorentz rotation of the tetrads. Indeed, this interpretation of spin connections as gauge fields of the Lorentz group (LGT) is not physically well-grounded in the Einstein-Cartan theory, since in this framework the spin connections do not include propagating degrees of freedom: they depend on tetrad fields and (algebraically), in the theory coupled with fermions, on the spin density too \cite{CianfrBook,Montani2,Montani4,Montani3,Nakia}.

A revised paradigm for this question is given by Poincar\'e gauge theory of gravity (PGT) \cite{Kibble,Sciama,Hehl0,hayashi,Hehl01,Hehl02,Obukhov}, where vierbein are identified with the gauge fields corresponding to the translational part of the Poincar\'e group. Hence, the gravitational interaction follows from the local extension of the invariance under Poincar\'e transformations.

The most general action containing up to second order derivatives of the tetrad fields and of the spin connections has been studied in \cite{Sezgin:1979zf,Hayashi:1980qp,Sezgin81,Hehl11} (a more restrictive action can be found in \cite{Kuhfuss} and the study on the renormalizability in \cite{Neville80}, see also \cite{Tseytlin:1981nu}). It has been pointed out how generically on a Minkowski background some ghosts and tachyon fields arise and a subset of theories has been identified in which these pathologies are avoided.   

In this work, we consider Fairchild's action \cite{Fairchild:1977wi}, which contains, besides the Einstein-Cartan term, the quadratic terms in the curvature tensor with close analogy with the free Yang-Mills terms for the Lorentz group. From the analysis in \cite{Sezgin:1979zf} (see also \cite{Hayashi:1980qp}), this theory is known to posses a spin-two pseudo-tensorial tachyon field. Our aim is to discuss in detail the phenomenology of this theory at the linear order with the aim to physically characterize at which level the pathology of the theory arises.

It is worth noting how we insist on considering, at least on a linear approximation, Fairchild's action \cite{Fairchild:1977wi}, 
because it constitutes the right formulation of gravity as the gauge field of
the Lorentz group, on the same footing of the standard Yang-Mills theories (when the fundamental Einstein-Cartan invariant, 
which vanishes for the Yang-Mills theory, is included). 

Despite the kinematic space of General Relativity closely resembles a Yang-Mills 
structure, two main shortcomings affect 
this representation \cite{Wein}: i) the spin-connection is, 
on shell, clearly dependent on the tetradic fields, {\it i.e.} the theory conserves a 
privileged role of the metric field representation \cite{Feynman:1996kb}; ii) the dynamics of the spin-connection is different from a
Yang-Mills theory, as a trivial consequence of the linearity of the theory Lagrangian in the gauge-curvature. 

The present approach simultaneously addresses these two shortcomings, 
restoring the full character of a Yang-Mills interaction in the gravitational physics. 

Clearly, in a more general Poincar\'e gauge field scenario, different scalar 
and tachyon free Lagrangian theories can be considered (see \cite{Sezgin:1979zf,Hayashi:1980qp,Sezgin81,Blagojevic:2013xpa}
) and they constitute interesting generalization of Einstein's gravity. However, the
present construction is the only one possessing the full features of a Lorentz gauge
theory, in view of a physically driven extension of the Riemannian 
geometry: including torsion in the space-time morphology allows to 
upgrade the spin-connection fields up to the role of real Yang-Mills potentials. 
The idea underlying the present letter is that such a special case deserves 
detailed investigations and the presence of a tachyon in itself is 
not a good reason to rule out, a priori, such a perspective. 

In this respect, we provide an interesting discussion about the possibility to constraint the coupling 
constant of the quadratic corrections in the Lagrangian, when considering 
a very precise measurement, like the muon gyromagnetic factor. 
However, the crucial point is 
whether or not the tachyon field interacts with ordinary matter fields, so yielding 
possible causality violation. The main merit of this study is to demonstrate how the 
backreaction of matter fields on the geometry provides a gravitational wave which in turn interacts with 
the tachyon field, leading to causality violation in the matter sector. 

In order to give a self-consistent presentation, we derive the equations of motion and we solve them in vacuum by linearizing the tetrad field and the connections around a flat space-time. We outline that torsion owns an intrinsically dynamical behaviour and, by splitting into irreducible components, we show that the propagating degrees of freedom are represented by the pseudo-scalar massive field and the tachyon field found in \cite{Sezgin:1979zf}. An interesting feature of this analysis is that the fields are massive and their masses are fixed by the coupling constant $\gamma$ for the Yang-Mills term, which is indeed the only free parameter of the model. Such Yang-Mills term is the only viable quadratic modification providing a non-trivial contribution to the equations of motion in the context of a Lorentz gauge theory.

The investigation on the behavior of spinor fields neglecting their backreaction on the geometry outlines that the tachyon field does not couple to spinors, thus the theory seems viable at the linear order. Hence, at this level only the pseudo-scalar field couples with fermions. In this case, the spin connections receive a contribution sourced by the fermion field, but the resulting interaction is very weak, since it is suppressed by the large value of the Planck mass. Such weak coupling between our model and the matter is also confirmed via the direct evaluation of the induced modification to the anomalous gyromagnetic factor for the $\mu$ particle. Then, we include backreaction and we outline how the energy momentum tensor of the spinor field generates a gravitational wave, coupling directly to the tachyon field.


\section{Lagrangian formulation}\label{III}

In the context of LGT, we consider the most general Lagrangian containing up to first derivatives in the fields, scalar under parity and reducing to General Relativity in the proper limit. According with these hypotheses, our Lagrangian formulation reads 
\begin{equation}\label{eq:AZIONEMODIFICATA}
S[e,\omega]=-\frac{1}{2\chi}\int d^{4}x\,e (R^{\quad ab}_{\mu\nu}\,e^{\mu}_{a}\,e^{\nu}_{b}+\gamma\,R^{\mu\nu}_{\quad ab}\,R_{\mu\nu}^{\quad ab}+\beta\,\eta^{\mu\nu\rho\sigma}\epsilon_{abcd}R_{\mu\nu}^{\quad ab}\,R_{\rho\sigma}^{\quad cd}),
\end{equation}
where $R_{\mu\nu}^{\quad ab}$ is the curvature tensor of the Riemamnn-Cartan spacetime
\begin{equation}
R_{\mu\nu}^{\quad ab}=\partial_{\mu}\omega_{\nu}^{\,\,ab}-\partial_{\nu}\omega_{\mu}^{\,\,ab}+\omega_{\mu}^{\,\,ac}\omega_{\nu c}^{\,\,\,\,\,\,b}-\omega_{\nu}^{\,\,ac}\omega_{\mu c}^{\,\,\,\,\,\,b}, 
\end{equation}
while $\chi=8\pi G$ and $\gamma$, by the analogy with Yang-Mills lagrangian, is a positive coupling constant (this requirement ensures that the theory is ghost-free). In what follows, $\beta$ will play no role, since it can be easily verified that it multiplies a topological term, and the action above reduces to Fairchild's \cite{Fairchild:1977wi}.

In the first order formalism, the spin connections $\omega^{ab}$ and the tetrads $e^{a}$ fields are treated as the basic independent variables. The equations of motion in vacuum follow from the variation of the action \eqref{eq:AZIONEMODIFICATA} with respect to these sets of variables, {\it i.e.}:
\begin{align}
&D_{\mu}\left[e\,e^{[\mu}_{a}\,e^{\nu]}_{b}+\,2\,\gamma \,e\,R^{\mu\nu}_{\,\,\,ab}\right]=0 \label{eq:VARIAZIONEOMEGA}\\
&R^{c}_{\rho}-\frac{1}{2}\left(R+\gamma \, R^{\quad ab}_{\mu \nu}\,R^{\mu \nu}_{\quad ab} \right) e^{c}_{\rho}+ 2\,\gamma\,R^{c}_{\,\,\,\,\nu\,ab}\,R^{\quad\nu ab}_{\rho}=0.\label{eq:VARIAZIONETETRAD}
\end{align}
The first term in \eqref{eq:VARIAZIONEOMEGA} is proportional to the torsion \cite{Hehl2,CianfrBook}, while the second corresponds to the equation of motion of a free Yang-Mills theory. It is natural to identify \eqref{eq:VARIAZIONETETRAD} as the form that Einstein equations of motion take in this formalism in absence of matter. 

\section{Linearised theory on Minkowski background}\label{IV}
The system of equations \eqref{eq:VARIAZIONEOMEGA} and \eqref{eq:VARIAZIONETETRAD} is non-linear both in tetrads and in spin connection fields. Since from \cite{Sezgin:1979zf} we know that there are pathologies in the propagator structure for torsion, let us now investigate the linearised field equations with respect to the torsion field on a Minkowski background, by fixing the tetrads as follows 
\begin{equation}\label{eq:gaugefix}
e^{a}_{\mu}(x)=\delta^{a}_{\mu}.
\end{equation}   
This particular choice provides an identification between space-time and internal indexes and a significant simplification of the dynamical problem: torsion-free spin connections vanish ($\bar{\omega}_{\mu}^{ab}=0$) and the full spin connection coincides with the contortion field components ($\omega_{\mu}^{ab}=K_{\mu}^{ab}$). Hence, the Riemann tensor can be written as
\begin{equation}
R_{\mu\nu}^{\quad ab}=2\,\partial_{[\mu}K_{\nu]}^{\,\,ab}+\circ(K^{2}).
\end{equation}
 The system of equations \eqref{eq:VARIAZIONEOMEGA} and \eqref{eq:VARIAZIONETETRAD} becomes at the linear order of approximation 
\begin{equation}\label{eq:sistema}
\begin{cases}
\Box\,K_{cab}-\partial_{c}\partial_{\mu}K^{\mu}_{\,\,\,ab}+\frac{1}{2\gamma\,}(-K^{\,\,\, d}_{d\,\,[a} \,\eta_{b]c}+K_{[ab]c})  =0\\
\partial^{c}(-K^{\,\,\, d}_{d\,\,[a} \,\eta_{b]c}+K_{[ab]c})   =0\\
\partial_{[\mu}K_{\rho]}^{\quad \mu b}=0,
\end{cases} 
\end{equation}
where $K_{cab}=e^{\mu}_{c}\, K_{\mu\,ab}=\delta^{\mu}_{c}\, K_{\mu\,ab}$, while the second equation is clearly just a consistency condition for the first one. 

In order to find the solutions of \eqref{eq:sistema}, we now split the contortion tensor into its irreducible components.

\subsection{Irreducible components of the contortion tensor}
Torsion can be decomposed into three irreducible tensors \cite{McCrea,Shapiro,Capozziello} (a multidimensional decomposition can be found in \cite{Shapiro2}). In this paragraph we decompose in a similar manner the contortion tensor and we show how this decomposition clarifies the nature of its propagating degrees of freedom.\\  
We can write the contortion tensor as follows:
\begin{equation}\label{decomposition}
K_{cab}=A_{cab}+v_{cab}+t_{cab}.
\end{equation}  
where we have isolated the totally-antisymmetric contortion part $A_{cab}$, {\it i.e.} 
\begin{equation}\label{eq:totallyanti}
A_{cab}=K_{[cab]}=-\frac{1}{6}\epsilon_{cabd}A^{d},
\end{equation}
corresponding to an axial-vector $A^{d}=\epsilon^{dcab}K_{cab}$ having four independent components, and the trace part $v_{cab}$ 
\begin{equation}\label{eq:trace}
V_{cab}=\frac{1}{3} \left(\eta_{ca}K_{b}-\eta_{cb}K_{a}\right),
\end{equation}
with the polar vector $K_{a}=K^{c}_{\,\,\,\,ca}$ having also four independent components. The last term in \eqref{decomposition} tensor is traceless and its totally-antisymmetric part vanishes, from which it follows that there are sixteen independent components in $\mathcal{T}_{cab}$. This reconciles the total number of independent components with that of the contortion tensor, which is twenty-four.

\subsection{Dynamic properties of the irreducible tensors}

Let us now investigate the implications of \eqref{eq:sistema} for each irreducible tensor in \eqref{decomposition}. As soon as $A_{abc}$ is concerned, let us note that the condition $\partial^{c}A_{cab}=0$ implies $\partial_{[\mu}A_{\rho]}^{\quad \mu b}=0$, thus we get from \eqref{eq:sistema} 
\begin{equation}\label{eq:sistema1}
\begin{cases}
\Box\,A_{cab}-\partial_{c}\partial^{d}A_{d\,ab}+\frac{1}{2\gamma\,}A_{cab}=0\\
\partial^{c}A_{cab}=0.\\
\end{cases}
\end{equation}
The second condition in \eqref{eq:sistema1} rewrites by means of \eqref{eq:totallyanti}
\begin{equation}\label{eq:Poinc}
\epsilon_{cabd}\partial^{[c}A^{d]}=0
\end{equation}
and if the spacetime manifold is simply-connected it implies that $A^{d}=\partial^{d}A$, for some pseudo-scalar field $A(x)$. The first condition becomes
\begin{equation}\label{eq:sistem1}
\Box\,A+\frac{1}{2\gamma\,}A=0,
\end{equation}
which is the Klein-Gordon equation for a field with mass $m=1/2\gamma$. Therefore, the totally-antisymmetric component of the contortion tensor carries one degree of freedom in the form of a massive pseudo-scalar field.\\
Similarly, from the second equation in \eqref{eq:sistema}, we get the following condition for $V_{cab}$ \eqref{eq:trace}
\begin{equation}\label{eq:second}
\partial^{c}\,V_{cab}=\partial_a K_b-\partial_b K_a=0,
\end{equation} 
the only possible solution in a simply-connected manifold being $K_{a}=\partial_{a}v$. The other equations become 
\begin{equation}\label{eq:sistem12}
\begin{cases}
\Box\,v+\frac{1}{2\gamma\,} v=0\\
\delta^{b}_{\rho} \Box v-2\partial_{\rho}\partial^{b}v=0
\end{cases} 
\end{equation}
and they admit only the trivial solution $v=0$ (this can be seen by multiplying the second condition times $\delta_{b}^{\rho}$). 
Finally, the system \eqref{eq:sistema} for the last part of the contortion tensor reduces to 
\begin{equation}\label{eq:sistem3}
\begin{cases}
\Box\,\, \mathcal{T}_{cab}-\frac{1}{4\gamma\,}\, \mathcal{T}_{cab} =0\\
\partial^{c}\,\mathcal{T}_{cab} =0\\
\partial^{c}\, \mathcal{T}_{abc}  =0.
\end{cases} 
\end{equation}
From the first equation we see how $\mathcal{T}_{cab}$ describes the propagation of a tachyon particle. The other conditions can be solved by fixing the frame in which the four-momentum $k_\mu=(0,0,0,1/\sqrt{4\gamma})$ and by requiring $\mathcal{T}_{cab}$ to vanish when one of the index 
$a,b,c=3$ (this can be easily seen in Fourier space). Given these conditions, the total number of independent components within $\mathcal{T}_{cab}$ is five.

The generality of our solutions can be verified by counting the physical degrees of freedom. Spin connections $\omega_{\mu}^{\,\,\,ab}$ have 24 components, but the six components $\omega_{0}^{\,\,\,ab}$ must be removed, because of their non-dynamical character (their time derivatives are not present in the action \eqref{eq:AZIONEMODIFICATA}). Moreover the condition \eqref{eq:VARIAZIONETETRAD}, together with Lorentz invariance, removes twelve additional components, so that the theory is eventually characterised by six physical degrees of freedom only. It is easy to check that the solutions we have found contains the correct number of physical degrees of freedom: one degree of freedom associated with the pseudo-scalar field $A$ and five degrees of freedom corresponding to $\mathcal{T}_{cab}$. \\
Therefore, the contortion tensor solving the equations of motion of the model reads
\begin{equation}\label{eq:solution}
K_{cab}=-\frac{1}{6}\epsilon_{cabd}\,\partial^{d}A + \mathcal{T}_{cab}.
\end{equation}
In the next section we will outline how only the pseudo-scalar field $A$ interacts with spinor fields, while the tachyon field decouples (at least classically), thus suggesting that un-physical interactions do not occur.

\section{Field equations in presence of spinors}\label{V}
In this section we investigate the role of spinor fields on the curved space-time whose dynamic is described by the action \eqref{eq:AZIONEMODIFICATA} (see \cite{Shirafuji:1979bm} for early studies on this subject).\\
The internal Lorentz gauge symmetry acts on spinor fields just like Yang-Mills gauge symmetries \cite{Hehl2, Utiyama} and the total action can be written as 
\begin{equation}\label{eq:AzioneSpin}
S=S_{g}[e,\omega]+S_{m}[e,\psi,D_{\mu}\psi],
\end{equation}
where the spinor action $S_m$ reads 
\begin{equation}\label{eq:DiracSimm}
S_{m}[e,\psi,D_{\mu}\psi]=\int dx^{4}\,e\,\bigg[\frac{i}{2}\left( \bar{\psi}\gamma^{\mu}\partial_{\mu}\psi- \left(\partial_{\mu}\bar{\psi}\right)\gamma^{\mu}\psi \right)
+\frac{1}{4}\epsilon_{cabd}\,\omega^{[c\,ab]}\bar{\psi}\gamma_{5}\gamma^{d}\psi-m\bar{\psi}\psi\bigg].
\end{equation}
It is worth noting that the action above contains an explicit coupling between spinor fields and spin connections, thus spinor enters the I Cartan equation and provides a nonvanishing contribution to torsion.   

\subsection{Torsion-spinor coupling}
We now consider the interaction between spin connections and spinors at the leading order of a perturbative expansion. Hence, we substitute the vacuum spin connections \eqref{eq:solution} into the Dirac action \eqref{eq:DiracSimm}. Since spinors couples only to the total antisymmetric part of the connection and the total antisymmetric part of the tachyon field by construction vanishes, the tachyon field $q_{cab}$ does not interact with them (at least at the leading order of the perturbative expansion). This is a relevant phenomenological issue, because it ensures physical viability to the considered quadratic modification of gravity, since the predicted tachyon field is indeed an isolated mode of the theory spectrum.

So the only contribution is given by the pseudo-scalar field \eqref{eq:sistem1} and the spinor lagrangian reads
\begin{equation}\label{eq:lagrIntMuon}
\mathcal{L}=\frac{i\hbar c}{2}\left[ \bar{\psi}\gamma^{\mu}\partial_{\mu}\psi- \left(\partial_{\mu}\bar{\psi}\right)\gamma^{\mu}\psi \right]-\frac{\hbar c}{4}(\partial_{a}A)\,\,\bar{\psi}\gamma^{a}\gamma^{5}\psi-\hbar c\,m\bar{\psi}\psi.
\end{equation}
The interaction term can be integrated by parts, so getting
\begin{equation}\label{eq:lagrIntMuon3}
\mathcal{L}=\frac{i\hbar c}{2}\left[ \bar{\psi}\gamma^{\mu}\partial_{\mu}\psi- \left(\partial_{\mu}\bar{\psi}\right)\gamma^{\mu}\psi \right]+\frac{i \hbar c}{2}m\,A\,(\,\bar{\psi}\gamma^{5}\psi)-\hbar c\,m\bar{\psi}\psi,
\end{equation}
where the following relation has been used 
\begin{equation}
\partial_{a}(\,\bar{\psi}\gamma^{a}\gamma^{5}\psi)= 2i m \bar{\psi}\gamma^{5}\psi.
\end{equation}
Let us now redefine $A$ as 
\begin{equation}
A\rightarrow\sqrt{\frac{6}{\chi}} \,A\,,
\end{equation}
such that it has the dimensionality of a scalar field (this can be seen from its kinetic term), while the coupling term with spinors rewrites 
\begin{equation}\label{eq:intWellDef}
\mathcal{L}_{int}=i\,g\,A\,\bar{\psi}\gamma^{5}\psi,\qquad g=\sqrt{\frac{\pi}{3}}\,\frac{m}{M_{p}} ,
\end{equation}
$M_p = \sqrt{\hbar c/\chi}$ being the Planck mass. Therefore, the coupling constant $g$ between spinors and the pseudo-scalar torsion component depends on the fermion mass. However, in view of the hierarchy between particle and Planck masses the value of $g$ is much smaller than the coupling constants of other interactions.

In order to estimate the possible phenomenological implications of our model, we evaluate the contribution given by the interaction with the pseudo-scalar field $A$ to the gyro-magnetic moment of a lepton \cite{deltaamu}, finding a displacement with respect to the standard value 
\begin{equation}\label{eq:deltaamu}
\Delta a=-\frac{g^{2}}{8\pi^{2}}\,\lambda^{2}\,\int_{0}^{1}\,dx\,\frac{x^{3}}{(1-x)(1-\lambda^{2}x)+\lambda^{2}x},
\end{equation}
where $\lambda=\frac{m}{M_{\Omega}}$ and $M_{A}=(2\gamma)^{-1/2}$ is the pseudo-scalar field mass. 
The maximum of $\Delta a$ is reached for $\lambda\rightarrow\infty$ and it reads
\begin{equation}\label{eq:limit1}
\left|\Delta a\right|= \frac{1}{2}  \frac{g^{2}}{8\pi^{2}},
\end{equation}
which is suppressed by the factor $g^2$. For instance, for a $\mu$ particle, $
g\approx 10^{-20}$ and the corresponding $\Delta a$ is several orders of magnitude below the experimental uncertainty \cite{PDG}.
Therefore, we do not expect any sensible deviation to the standard particle physics phenomenology coming from our model. 

\subsection{Tachyon-spinor interaction} 
The backreaction of the spinor field on the geometry can be investigated by perturbing the tetrad, {\it i.e.} $e^a_\mu=\delta^a_\mu+h^a_\mu$. $h^a_\mu$ 
describe the gravitational wave generated the spinor field and using the gauge in which $\partial^\mu h^a_{\mu}=0$, the equations of motion \eqref{eq:VARIAZIONEOMEGA} and \eqref{eq:VARIAZIONETETRAD} rewrite
\begin{align}
&-2\partial_{[a}\Box h_{b]c} + \Box K_{cab}-\partial_{c}\partial^{d} K_{dab}+\frac{1}{2\gamma}\left[-K^{d}_{\,\,\,d[a}\eta_{b]c}+K_{[ab]c}\right]=\frac{\chi}{4\gamma}\epsilon_{cabd}\,\bar{\psi}\gamma_{5}\gamma^{d}\psi\label{eq:mm}\\
&-\Box h_{\mu}^{c} - \partial_{\mu}\partial^{c}h+ 2\partial_{[a}K_{\mu]}^{\,\,ac}=\frac{i}{2}\left[\bar{\psi}\gamma^{c}\partial_{\mu}\psi - (\partial_{\mu}\bar{\psi})\gamma^{c}\psi\right]+\frac{i}{4}\, \omega_{\mu ab}\,\bar{\psi}\gamma^{[c}\gamma^{a}\gamma^{b]}\psi -\frac{1}{2}m\bar{\psi}\psi\delta^{c}_{\mu}.\label{eq:Eins}
\end{align}
$h$ being the trace of $h^a_\mu$, namely $h=h^a_\mu \,\delta^\mu_a$. We can still split the contortion components as in \eqref{decomposition}. It turns out from \eqref{eq:mm} that spinor axial current couples only to the pseudo-scalar field
\begin{equation}
\Box\Omega_{cab}+\frac{1}{2\gamma}\Omega_{cab}=\frac{\chi}{4\gamma}\epsilon_{cabd}\bar{\psi}\gamma_{5}\gamma^{d}\psi,
\end{equation}
while for the other irreducible contortion components the following equations holds
\begin{align}
&\Box \phi+\frac{1}{4\gamma}\phi= -\Box h\\
&\Box q_{cab}-\frac{1}{2\gamma}q_{cab} =2\partial_{[a}\Box h_{b]c} +\eta_{c[a}\partial_{b]}\Box h\label{fint}\,.
\end{align}
It is worth noting from \eqref{fint} that the tachyon field is sourced by the gravitational wave. Moreover, by rewriting \eqref{eq:Eins} in terms of irreducible contortion components one ends up with the following equation 
\begin{equation}
\Box h_{\mu}^{a} - \partial_{\mu}\partial^{a}h=\frac{i}{2}e\left[\bar{\psi}\gamma^{a}\partial_{\mu}\psi - (\partial_{\mu}\bar{\psi})\gamma^{a}\psi\right]-m\bar{\psi}\psi\delta^{a}_{\mu} + \partial_{\mu}\partial^a\phi-\frac{1}{2}\delta^a_\mu\,\frac{1}{4\gamma}\phi
\end{equation}
which shows how the gravitational wave is sourced by the spinor field. Henceforth, the spinor field generates a gravitational wave, which couples directly with the tachyon field. This results in an effective interaction between fermions and the tachyon, which provide causality violation in the matter sector.

\section{Conclusions}\label{VII}

In this work we considered a propagating torsion theory, obtained by adding to the Einstein-Cartan term a quadratic contribution in the curvature, which resembles a Yang-Mills action for the spin connection. We analyzed classical equations of motion and we solved them on a Minkowski background in the linearised limit. Torsion is generically nonvanishing also in vacuum and it carries five degrees of freedom, described by a pseudo-scalar field and a tachyon particle. Ignoring matter backreaction on the geometry, the spinor fields see only the pseudo-scalar component via a coupling constant suppressed by the ratio between the mass of the spinor field and Planck mass. As soon as backreaction is included, spinors generate gravitational waves, which act as sources for the tachyon field.

The main contribution of the present analysis consists of fixing how the tachyon field couples with ordinary matter. 
Indeed, such an information came out from a detailed study of the field equations. Up to linear theory on a Minkowski 
space-time, we got the surprising result that the tachyon degree of freedom remains an isolated mode 
and it couples only when a gravitational wave is generated by the spinor 
energy-momentum tensor. Despite its extremely small amplitude, the tachyon-gravitational 
wave coupling suggests causality violation in the matter sector since it provides an effective interaction between the tachyon and 
spinor field, for which causality violation is immediately inferred 
in terms of the interaction cross section. 

The present study says a clear definitive word on the non-viability of the considered 
theory, which is of particular interest because it represents the natural extension of 
the Einstein-Hilbert action for gravity towards the construction of a non-Abelian 
gauge theory of the Lorentz group. 

\acknowledgments

FC is supported by funds provided by the National Science Center under the agreement
DEC-2011/02/A/ST2/00294.

\end{document}